\documentstyle[prl,aps,twocolumn,floats,psfig]{revtex}

\renewcommand{\vec}[1]{{\bf #1}}

\newcommand{\be}{\begin{equation}}
\newcommand{\ee}{\end{equation}}
\newcommand{\ber}{\begin{eqnarray}}
\newcommand{\eer}{\end{eqnarray}}

\begin{document}
\draft

\title{Charge ordering and hopping in a triangular array of quantum dots}
\author{L. S. Levitov$^1$, Boris Kozinsky}
\address{Center for Materials Science \& Engineering, 
Physics Department, Massachusetts Institute of Technology\\
$^1$ Also, Department of Condensed Matter Physics, The Weizmann
  Institute of Science, Rehovot 76100, Israel}

\date{\today}

%\begin{document}

\maketitle

%\thanks{Physics Department, Massachusetts Institute of Technology}

  \begin{abstract}
We demonstrate a mapping between the problem of charge ordering in 
a triangular array of quantum dots and a frustrated Ising
spin model. Charge correlation in 
the low temperature state is characterized by an intrinsic
height field order parameter. Different ground states are possible
in the system, with a rich phase diagram. We show that 
electronic hopping transport is sensitive to the properties of
the ground state, and describe the singularities of hopping conductivity
at the freezing into an ordered state. 
  \end{abstract}
\bigskip 
%\pacs{PACS: 73.40.Gk, 73.40.Hm}

%%%%%%%%%%%%%%%%%%%%%%%%%%%%%%%%%%%%%%%%%%%%%%%%%%%%%%%%%%%%%%%%%%%%%%%%%%
%
%
%
%%%%%%%%%%%%%%%%%%%%%%%%%%%%%%%%%%%%%%%%%%%%%%%%%%%%%%%%%%%%%%%%%%%%%%%%%%

Ordering and phase transitions in artificial
structures such as Josephson junction arrays\cite{Josephson-arrays} 
and arrays of quantum dots\cite{dot-arrays} 
have a number of interesting properties. One attractive feature of these systems
is the control on the Hamiltonian by the system design. 
Also, the experimental techniques available for 
probing magnetic flux or charge ordering, 
such as electrical transport measurements and scanning probes,
are more diverse and flexible than those conventionally used 
to study magnetic or structural ordering in solids. 
There has been a lot of theoretical\cite{Josephson-arrays-theory}
and experimental\cite{Josephson-arrays} studies of 
phase transitions and collective phenomena 
in Josephson arrays, and also some work on the quantum 
dot arrays\cite{dot-arrays-experiment,dot-arrays-theory}. 

There is an apparent similarity between the Josephson 
and the quantum dot arrays problems, because
the former problem can be mapped using a duality transformation 
on the problem of charges on a dual lattice. These charges 
interact via the $D=2$ Coulomb logarithmic potential, and can exhibit
a Kosterlitz-Thouless transition, as well as other interesting 
first and second order transitions\cite{Josephson-arrays-theory}.
In the quantum dots arrays, charges are coupled via the $D=3$ Coulomb
$1/r$ potential, which can be also partially screened by ground electrodes or gates. 
In terms of the interaction range, the quantum dot array system 
is intermediate between the Josephson array problem and the 
lattice gas problems with short range interactions, such as the Ising 
model and its varieties. From that point of view, an outstanding question is
what are the new physics aspects of this problem. 

A very interesting system fabricated recently\cite{Bawendi} is based on
nanocrystallite quantum dots that can be produced with high 
reproducibility, of diameters $\sim 15-100$\AA
tunable during synthesis, with a narrow size 
distribution ($<5\%$ rms). These dots can be forced to assemble into
ordered three-dimensional closely packed colloidal crystals\cite{Bawendi},
with the structure of stacked two-dimensional triangular lattices. 
Due to higher flexibility and structural control, these systems
are expected to be good for studying effects inaccessible in 
the more traditional self-assembled quantum dot arrays fabricated 
using epitaxial growth techniques. 
In particular, the high charging energy of 
nanocrystallite dots, in the room temperature range, and the triangular lattice 
geometry of the dot arrays\cite{Bawendi} 
are very interesting from the point of view of
exploring novel kinds of charge ordering. 

Motivated
by recent attempts to bring charge carriers into these structures using gates,
and to measure their transport properties\cite{Kastner}, 
in this article we study charge transport in the classical
Coulomb plasma on a triangular array of quantum dots. 
We assume that the dots can be charged by an external gate and
that conducting electrons or holes can tunnel between neighboring dots. 

For drawing a connection to the better studied spin systems, 
it is convenient to map this problem on 
the classical Ising antiferromagnet on a triangular lattice ($\triangle$IAFM).
Without loss of generality, we
consider the case when the occupancy of the dots is either 1 or 0, 
and interpret these occupancies as the spin ``up'' and ``down'' states. 
In this language, the gate voltage is represented by an external magnetic field
coupled to the spins. Also, more generally, 
any spatially varying electrostatic field 
is mapped on a spatially varying magnetic field in the spin problem. 
For example, the electrostatic field due to, 
e.g., charged defects, corresponds to the spin problem with a random field. 
The electron hopping (tunneling) between the dots corresponds to spin exchange 
transport. The long-range Coulomb interaction between charges 
gives rise to a long-range spin-spin coupling, which leads to somewhat different
physical properties than the short-range exchange interaction conventional for
the spin problems. 

Another difference between the charge and spin problems 
is that, due to charge conservation, there is no analog of spin flips.
This certainly has no effect on the equilibrium statistical
mechanics, because the spin ensemble with fixed total spin is 
statistically equivalent to the grand canonical ensemble. 
However, this is known to be important in the dynamical problem\cite{Hohenberg-Halperin}.
The two corresponding types of dynamics are the spin conserving Kawasaki dynamics\cite{Kawasaki} 
and the spin non-conserving Glauber dynamics\cite{Glauber},
respectively. In the simulation described below we use the Kawasaki dynamics, 
involving spin exchange processes on neighboring sites.

The mapping between the charge and spin problems is of interest
because of the following. If the interaction between the dots were of a purely 
nearest neighbor type,
the problem could have been exactly mapped on the 
$\triangle$IAFM problem, which is exactly solvable\cite{IsingAFM}.
The $\triangle$IAFM problem is known to
have an infinitely degenerate ground state with an intrinsic ``solid-on-solid''
structure described by the so-called ``height field''\cite{Blote} (see below). The
height field represents the correlations of occupancy of neighboring sites, and can be
thought of in terms of an embedding of the structure into a
three-dimensional space. The higher-dimensional representations
of correlations in solids have also been found useful in a variety of frustrated
spin problems\cite{q-magnets}.
The ordering of electrons in triangular arrays of quantum dots,
because of Coulomb coupling giving rise to 
a repulsive nearest-neighbor interaction, must be similar to that of
the $\triangle$IAFM ground states. It represents, however, a new
physical system in which the height field will be
strongly coupled to electric currents, which can make electronic transport 
properties very interesting. 

The Hamiltonian of the electrons on the quantum dots is given by 
${\cal H}_{\rm charge}+{\cal H}_{\rm tunnel}+{\cal H}_{\rm spin}$,
where ${\cal H}_{\rm charge}$ describes Coulomb interaction between charges
$q_i=0,1$ on the dots and coupling to the background disorder potential $\phi(r)$
and to the gate potential $V_{\rm g}$:
  \begin{equation}\label{E_tot}
{\cal H}_{\rm charge}=\frac{1}{2}\sum\limits_{i,j}V(\vec r_{ij})q_i q_j
+\sum\limits_{\vec r_i} (V_{\rm g}+\phi(\vec r_i))q_i
  \end{equation}
The position vectors $\vec r_i$ run over a triangular lattice with the lattice constant $a$,
and $\vec r_{ij}=\vec r_i-\vec r_j$.
The interaction $V$ accounts for screening by the gate:
  \begin{equation}\label{interaction}
V(\vec r_{ij}\ne 0)=\frac{e^2}{\epsilon|\vec r_{ij}|}
-\frac{e^2}{\epsilon \sqrt{(\vec r_{ij})^2+(2d)^2}}\ ,
  \end{equation}
Here $\epsilon$ is the dielectric constant
of the substrate\footnote{
   In the case of spatially varying $\epsilon$ the interactions can be 
more complicated. For example, if the array of dots is placed over a semiconductor
substrate, one has to replace $\epsilon\to (\epsilon+1)/2$ in 
the expression (\ref{interaction}).
   }, and $d$ is the distance to the gate plate.
The single dot charging energy $\frac{1}{2}V(0)=e^2/2C$ is assumed to be 
high enough to maintain no more than single occupancy.

Electron tunneling between neighboring dots is described by ${\cal H}_{\rm
tunnel}$. We assume that the tunneling is incoherent, i.e., 
is assisted by some energy relaxation mechanism, 
such as phonons. Below we consider stochastic
dynamics in which charges can hop between neighboring dots with
probabilities depending on the potentials of the dots and on
temperature. Also, we consider only charge states and ignore all
effects of electron spin described by ${\cal H}_{\rm spin}$, such as exchange, spin
ordering, etc.

\begin{figure}
\centerline{\psfig{file=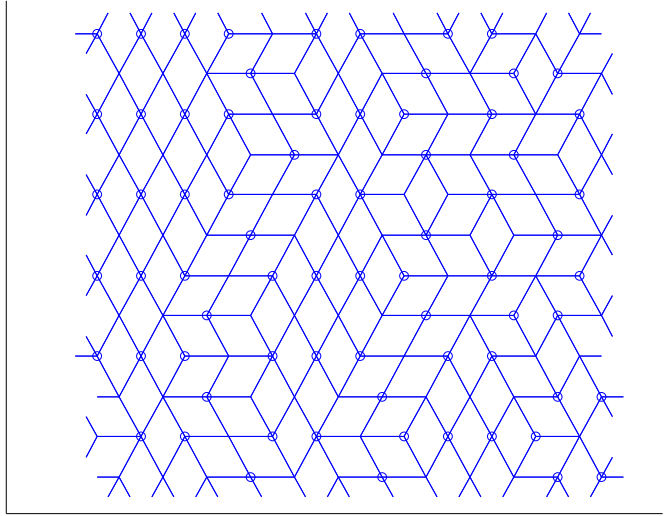,width=3.5in}}
\vspace{0.5cm}
 \caption[]{ Typical charge configuration obtained in a Monte Carlo 
        simulation for the filling fraction $n=1/2$. Pairing of the
        triangles is revealed by erasing all frustrated bonds connecting sites with 
        equal occupancy. 
    }
 \label{n=72}
\end{figure}

To lay out the framework for discussing ordering in the charge problem, let us 
review here the main results for the $\triangle$IAFM problem using 
the charge problem language (and assuming nearest neighbor interactions). 
The ground state of this system
has a large degeneracy\cite{IsingAFM}, which can be understood as follows. For
each triangular plaquette at least one bond is frustrated, 
for any occupancy pattern. To minimize the energy of the nearest neighbor 
interactions, i.e., to reduce the number of frustrated bonds,
it is favorable to arrange charges so that the triangles combine in
pairs in such a way that each pair of triangles shares a common frustrated bond.
The pairing of triangles can be described by partitioning
the structure into rhombuses. Given a charge configuration, 
the corresponding rhombuses pattern can be visualized by
erazing all frustrated bonds, as shown in Fig.\ref{n=72}.

%%%%%%%%%%%%%%%%%%%%%%%%%%%%%%%%%%%%%%%%%%%%%%%%%%%%%%%%%%%%%%%%%%%%
%%%%%%%%%%%%%%%%%%%%%%%%%%%%%%%%%%%%%%%%%%%%%%%%%%%%%%%%%%%%%%%%%%%%

Conversely, given a configuration of rhombuses, one can
reconstruct the arrangement of charges in a unique way, up to an
overall sign change. The use of the representation involving rhombuses is
that it leads to the notion of a height field. Any
configuration of rhombuses can be thought of in terms of a
projection of a faceted surface in a 3D cubic lattice along the
$(111)$ direction. This surface defines lifting of the 2D
configuration in the 3D cubic lattice space, 
i.e., an integer-valued height field. The number of
different surfaces that project onto the domain of area ${\cal A}$ is of
the oder $e^{w{\cal A}}$, where $w$ is a constant equal to the entropy 
of the ground state manifold per plaquette.

\begin{figure}
\centerline{\psfig{file=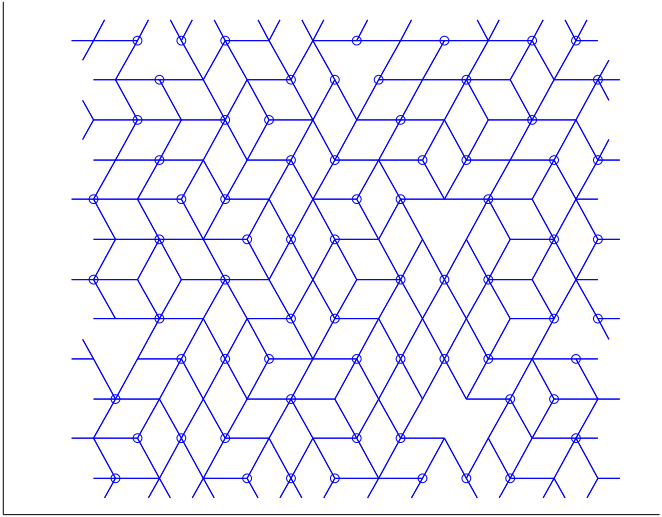,width=3.5in}}
\vspace{0.5cm}
 \caption[]{ Typical charge configuration for the filling fraction $n=1/2$. In this case,
             the temperature is warmer than in Fig.\ref{n=72}, and there are several
             defects (``screw dislocations'') present in the system.
    }
 \label{n=72dis}
\end{figure}

At a finite temperature, the charge ordering with respect to 
the height field, i.e., to the pairing of triangles, may have some
defects (see Fig.\ref{n=72dis}). An elementary defect is represented by an isolated
unpaired triangle surrounded by rhombuses, i.e., by paired triangles.
This defect has a topological character similar to a screw dislocation,
because it leads to an ambiguity in the height field. This ambiguity is readily 
seen in Fig.\ref{n=72dis},
where by continuing the height field around an unpaired triangle
one finds a discrete change in it upon returning to the starting point.
 
%%%%%%%%%%%%%%%%%%%%%%%%%%%%%%%%%%%%%%%%%%%%%%%%%%%%%%%%%%%%%%%%%%%%
%%%%%%%%%%%%%%%%%%%%%%%%%%%%%%%%%%%%%%%%%%%%%%%%%%%%%%%%%%%%%%%%%%%%

There is no finite temperature phase transition in the $\triangle$IAFM problem because
of the topological defects present at a finite concentration at any temperature. 
However, since the fugacity of a defect scales as $e^{-V(a)/T}$, and the 
defect concentration scales with the fugacity, the system of defects
is becoming very dilute as $T\to 0$. As a result, the $T=0$ state is ordered, and 
belongs to the degenerate manifold of ground states without defects, 
i.e., is characterized by a globally defined height variable. 
Also, since the distance between defects diverges exponentially as $T\to 0$, there
is a large correlation length for the $T=0$ ordering even when $T$ is finite. 
This situation is described in the literature on the $\triangle$IAFM problem 
as a $T=0$ critical point (see recent article \cite{chakraborty} and references therein). 

In this work, we study the charge system with the interaction (\ref{interaction}) 
by a Monte Carlo (MC) simulation of electron hopping in equilibrium, as well
as in the presence of an external electric field. 
We find that, although many aspects of the $\triangle$IAFM physics
are robust, the long-range character of the interaction (\ref{interaction})
makes the charge problem different from the $\triangle$IAFM problem in several ways. 

\begin{figure}
\centerline{\psfig{file=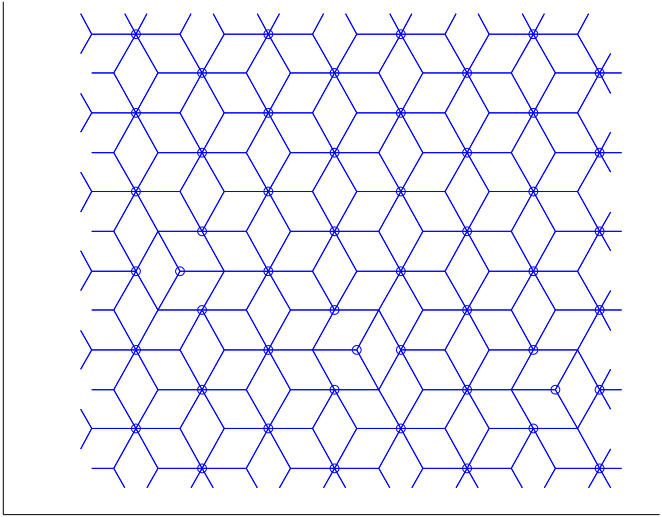,width=3.5in}}
\vspace{0.5cm}
 \caption[]{ Typical Monte Carlo charge configuration for $n=1/3+\epsilon$. 
             There are three excess charges in the system ($\epsilon=3/144$)
           hopping over the honeycomb network of unoccupied sites in the  
           commensurate $\sqrt{3}\times\sqrt{3}$ state. The excess carriers,
           dilute at $\epsilon\ll1$, are moving nearly independently 
           on the background of the frozen $\sqrt{3}\times\sqrt{3}$ state.
           
    }
 \label{n=51}
\end{figure}

The states undergoing the MC dynamics are all charge configurations with no more 
than single occupancy ($q_i=0,1$) on a square patch $N\times N$ of a triangular array 
($N=12$ in Figs.\ref{n=72},\ref{n=72dis},\ref{n=51}). 
Periodic boundary conditions are imposed by defining the energy (\ref{E_tot}) 
using the $N\times N$ charge configuration extended
periodically in the entire plane. Also, we allow charge hopping across the patch boundary,
so that the charge disappearing on one side of the patch reappears on the opposite side,
consistent with the periodicity condition. 

The stochastic MC dynamics is defined by letting electrons hop on 
unoccupied neighboring sites with probabilities given by Boltzmann weights:
  \be\label{probabilities}
W_{i\to j}/W_{i\to i}=e^{(\Phi_i-\Phi_j)/kT}\ ,\ 
W_{i\to j}+W_{i\to i}=1\ ,
  \ee
where 
  \be  
\Phi_i=\sum\limits_{r_j\ne r_i}V(\vec r_{ij})q_j + V_{\rm g} + \phi(\vec r_i)
\ .
  \ee
To reach an equilibrium at a low temperature, we take the usual 
precautions by running the MC dynamics first at some high temperature, 
and then gradually decreasing the temperature to the desired value. 

%Most of 
All the work reported below was done on the system without disorder, $\phi(\vec r_i)=0$.
The distance to the gate which controls the range of the interaction (\ref{interaction})
was chosen to be $d=2$.
%\ftn{  {\bf Borya:} Is it right?   }

The properties of the system depend on the charge filling fraction, i.e., on the mean 
occupancy $n=\sum q_i/N^2$, which is conserved in the MC dynamics. 
(The $\triangle$IAFM in the absence of external magnetic field corresponds to $n=1/2$.) 
We find that at low temperature the equilibrium states with $1/3\le n\le 2/3$ 
are very well described by pairing of the triangles, as illustrated in 
Figs.\ref{n=72},\ref{n=51}. The defects with respect
to this height field ordering have a very small concentration, if present at all. 
The short-range ordering in the charge problem turns out to be the same as 
in the $\triangle$IAFM. Qualitatively, this similarity is explained by 
a relatively higher magnitude of the nearest neighbor coupling (\ref{interaction})
compared to the coupling at larger distances. 

Like in the $\triangle$IAFM, we observe topological defects. 
They are present at warmer temperatures (see Fig.\ref{n=72dis}) and quickly 
freeze out at colder temperatures (see Fig.\ref{n=72}). The disappearance
of the defects, because of their topological character, is possible
only via annihilation of the opposite sign defects. 
An example of such a process can be seen in the lower right corner of Fig.\ref{n=72dis}. 
The defects in the charge problem, besides carrying topological charge, 
can carry an electric charge. Two defects of opposite electric charge can be found 
at the top and on the left of Fig.\ref{n=72dis}. 

%%%%%%%%%%%%%%%%%%%%%%%%%%%%%%%%%%%%%%%%%%%%%%%%%%%%%%%%%%%%%%%%%%%%
%%%%%%%%%%%%%%%%%%%%%%%%%%%%%%%%%%%%%%%%%%%%%%%%%%%%%%%%%%%%%%%%%%%%

Expectedly, besides the robust features, such as the height field and 
the topological defects, there are certain non-robust aspects of the problem. 
The two most interesting issues that we discuss here are related with 
lifting of the degeneracy of the $\triangle$IAFM ground state, and with 
the instability of the $T=0$ critical point, both arising due to 
long-range coupling in the charge problem.

As we already have mentioned, the ground state
manifold of the $\triangle$IAFM is fully degenerate only for purely nearest neighbor 
interactions. The degeneracy is lifted even by a weak 
non-nearest-neighbor interaction\cite{AFM-ferro-NNN,higher-spin-AFM}.
For instance, for our problem with the interaction (\ref{interaction}) at $n=1/2$,
the favored ground states have the form of stripes, spaced by $\sqrt{3}$
in the lattice constraint units, which corresponds to electrons filling every other lattice row. 
Because there are three possible orientations of the stripes, the system cooled
in the $n=1/2$ state freezes in a state characterized by domains
of the three types. One of such domains of stripes can be seen in the upper left part of 
Fig.\ref{n=72}. Upon long 
annealing, the domains somewhat grow and occasionally coalesce. 
However, we were not able to determine whether the system always reaches
a unique ground state, or remains in a polycrystalline state of intertwining domains. 

This can be contrasted with the behavior at $n=1/3$, where the ground state 
is a $\sqrt{3}\times \sqrt{3}$ triangular lattice, 
corresponding to electrons filling one of the three
sublattices of the triangular lattice (see Fig.\ref{n=51}). 
The rotational symmetry of this state is the same as that of the underlying 
lattice. Because there is a (triple) translational degeneracy of the ground state, 
but no orientational degeneracy, the system always forms a perfect triangular structure 
upon cooling, without domains. For the densities close to $1/3$, the 
ground state contains $|n-1/3|$ charged defects, vacancies for $n<1/3$
and interstitials for $n>1/3$, formed on the background of the otherwise perfect
$\sqrt{3}\times \sqrt{3}$ structure. At $n>1/3$, the interstitials 
are moving over the honeycomb network of empty sites, as illustrated 
in Fig.\ref{n=51}.

%%%%%%%%%%%%%%%%%%%%%%%%%%%%%%%%%%%%%%%%%%%%%%%%%%%%%%%%%%%%%%%%%%%%
\begin{figure}
\centerline{\psfig{file=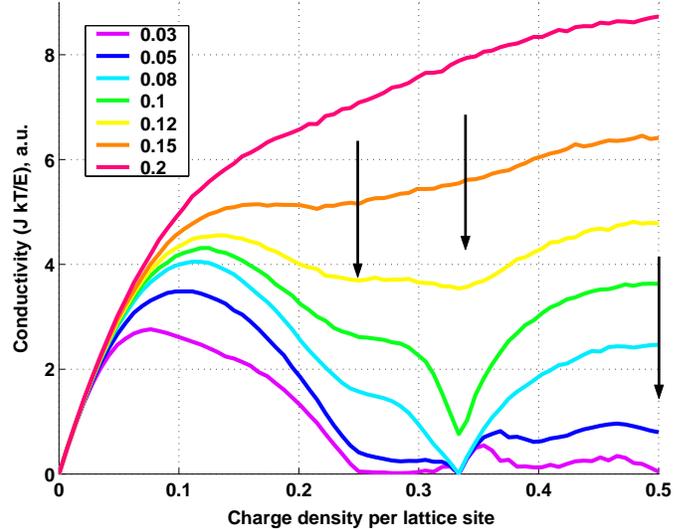,width=3.5in}}
\vspace{0.5cm}
 \caption[]{Conductivity versus electron density for several temperatures,
given in the units of $e^2/\epsilon a$. 
The density $n$ varies from $0$ (uncharged) to $1$ (fully charged),
however, because of the $n \leftrightarrow 1-n$ symmetry, only the 
interval $0\le n\le 1/2$ is displayed. 

Arrows mark the features corresponding to the phase transitions 
at $n=1/3, 1/2, 1/4$.
    }
 \label{T-n-plot}
\end{figure}
%%%%%%%%%%%%%%%%%%%%%%%%%%%%%%%%%%%%%%%%%%%%%%%%%%%%%%%%%%%%%%%%%%%%

Before discussing the issue of finite $T$ versus $T=0$ phase 
transitions, let us explain how we use the MC simulation to
find electrical conductivity. It is straightforward to add an external 
electric field $\vec E$ to the MC algorithm\cite{gas-dynamics}. For that, one can simply
modify the expressions (\ref{probabilities}) for the hopping probabilities
by adding the field potential difference $\vec E\cdot(\vec r_{ij})$ 
between the two sites $\vec r_i$ and $\vec r_j$. In doing this, one has
to respect the periodic boundary condition, which amounts to
taking the shortest distance between the sites $\vec r_i$ and $\vec r_j$ 
on the torus. Then the charges are statistically biased to 
hop along $\vec E$, which gives rise to a finite average current $\vec J$. 

The current $\vec J$ is Ohmic at small $\vec E$, and saturates at 
$|\vec E|a\ge {\rm min}(V(a)n^{1/2},\ kT)$, where $a(n)=n^{-1/2}$ is 
the inter-electron spacing. Even though only the Ohmic regime $J\sim\vec E$
is of a practical interest, it is useful to maintain
the field $\vec E$ used in the MC simulation on the level of
not less than few times below the nonlinearity offset field, to 
minimize the effect of statistical fluctuations on the time averaging of 
the current $\vec J$. 

The dependence of the electric current on the density $n$ is shown 
in Fig.\ref{T-n-plot} for several temperatures. Because of the electron--hole symmetry
of the system, the conductivity problem is invariant under the transformation 
$n\leftrightarrow 1-n$, and thus only the interval $0\le n\le 1/2$ can be considered. 
At high temperatures $kT\ge e^2/\epsilon a(n)$, the Coulomb interaction is 
not important, and one can evaluate the current by considering electrons freely 
hopping over lattice sites subject only to the single occupancy constraint. 
The current $\vec J$ in this case goes as 
  \be\label{J_free}
\vec J=\frac{3}{4} n(1-n)\, \frac{\vec E}{kT}\ ,
  \ee
where the factor $n(1-n)$ in (\ref{J_free}) gives the probability 
that a particular link connects
two sites of different occupancy, so that hopping along this link can occur,
whereas the factor $(\vec E/kT)$ in (\ref{J_free}) is the hopping probability 
bias due to the weak field
$|\vec E|\ll kT$. The constant factor $3/4$ in (\ref{J_free}) is given 
by the coordination number of the lattice (equal to 6) divided by 8. 

The parabolic $n(1-n)$ dependence of the current (\ref{J_free}) is clearly
reproduced by the highest temperature curve in Fig.\ref{T-n-plot}. In Fig.\ref{T-n-plot}
the inverse temperature factor of the expression (\ref{J_free}) is eliminated by rescaling 
the current by $E/kT$. 
The rescaled current $J(kT/E)$ in Fig.\ref{T-n-plot} 
is practically temperature independent 
at small densities $n$, in agreement with the result (\ref{J_free}). 
Regarding the $1/kT$ rescaling, note that in a real system the frequency
of electron attempts to hop is a function of temperature, because
hopping is assisted by energy relaxation processes, such as phonons.
In this case, our MC result for the current $\vec J$ has to be multiplied by some function
$A(T)$, which will be of a power law form $T^\alpha$ for the phonon absorption rate
in the case of a phonon-assisted hopping.

At the temperatures of the order and smaller than $e^2/\epsilon a(n)$, 
the current becomes suppressed due to charge correlations reducing 
the frequency of hopping. The onset of this suppression takes place
at temperatures of order of the interaction between
neighboring electrons, $kT\simeq 0.2 V(a)n^{1/2}$. Note that the onset temperature 
is lower for smaller density,
in agreement with the plots in Fig.\ref{T-n-plot}.

%%%%%%%%%%%%%%%%%%%%%%%%%%%%%%%%%%%%%%%%%%%%%%%%%%%%%%%%%%%%%%%%%%%%
\begin{figure}
\centerline{\psfig{file=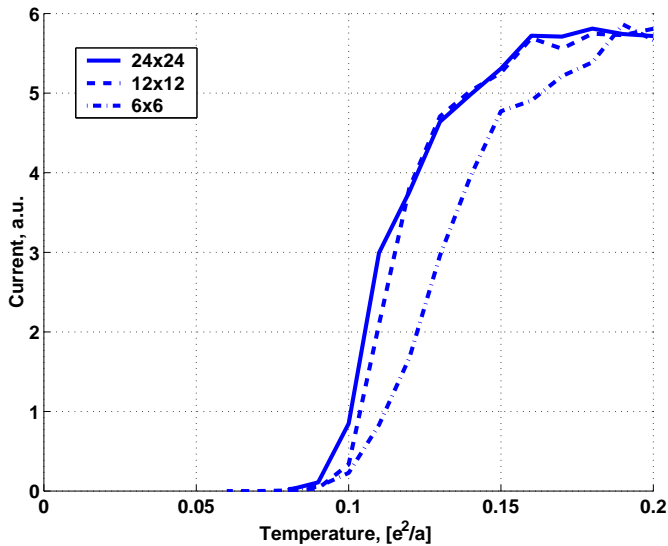,width=3.5in}}
\vspace{0.5cm}
 \caption[]{Conductivity versus temperature for 
density $n=1/3$. The three curves are obtained for 
three different system sizes.
    }
 \label{transition}
\end{figure}
%%%%%%%%%%%%%%%%%%%%%%%%%%%%%%%%%%%%%%%%%%%%%%%%%%%%%%%%%%%%%%%%%%%%

The electrical conductivity is a sensitive probe of freezing transitions,
at which the system of interacting charges locks collectively into a particular 
ordered or disordered state, and the conductivity vanishes. 
Whether the freezing occurs as a finite $T$ or a $T=0$ phase transition 
is related to the degree of degeneracy of the ground state. The situation 
appears to be very different at different densities $n$. 
For simple rational $n=1/3,\ 1/2,\ 2/3$, and alike, characterized by the ground state
unique up to discrete symmetries, there is a well defined freezing temperature. 
This is illustrated in Fig.\ref{transition}, where electrical conductivity 
of the $n=1/3$ state is plotted versus temperature. 
The abrupt drop of the conductivity at $kT\approx 0.09 e^2/\epsilon a$, where $a$ is the
lattice constant, indicates a sharp freezing transition. To make sure that
this is not a finite size effect, we show the conductivity curves for systems
of three different sizes, $6\times6$, $12\times12$, and $24\times24$. 

On the other hand, away from the specific densities with simple ground states, 
the freezing appears to be very gradual, and in the temperature 
range we explored in Fig.\ref{T-n-plot} there has been no evidence 
of a sharp transition. It may be that the ordering actually takes place at much smaller
temperatures than the characteristic interaction, the situation not uncommon 
in frustrated systems. Also, it may be that at incommensurate densities 
the state remains disordered down to $T=0$. From our observations, 
the latter seems to be a more likely scenario. We expect that upon cooling the charges freeze 
into a quasirandom state determined by the cooling history. In that case, this system
represents an electronic glass that exists in the absence of external disorder. 

The nature of the ground state in this case is unclear. To list several options, it may be that
the system forms a polycrystal consisting of intertwining domains, like it does at $n=1/2$,
or that the state represents a distorted incommensurate charge density wave, or that
it is a genuine glass. Studying this would require 
enhancing the MC algorithm to make it capable of treating the slow dynamics 
of annealing at low temperatures. 

In conclusion, correlations of charges in the triangular arrays are described by 
an intrinsic order parameter, the height field, similar to the $\triangle$IAFM problem. 
The charge--spin mapping shows that various interesting phenomena arising in
frustrated spin systems can be studied in charge systems, for which
more powerful experimental techniques are available. 
The type of order in the ground state depends on the charge filling density.
Electron hopping conductivity is sensitive to charge ordering and can be used as 
a probe of the nature of the ordered state. Conductivity couples to the height field, and
thus one can expect novel effects in electronic transport properties which have 
no analog in spin systems with the same geometry.

\acknowledgements
   We are very grateful to Marc Kastner, Moungi Bawendi, and Nicole Morgan for drawing our
attention to this problem, as well as for many useful discussions and for sharing 
with us their unpublished experimental results. This work is supported by 
the NSF Award 67436000IRG.

\end{document}